\documentclass[twocolumn,trackchanges]{aastex62}
\usepackage{amsmath}
\usepackage{graphicx}
\usepackage{amsfonts}
\usepackage{natbib}
\usepackage{color}
\usepackage{epstopdf}
\usepackage{graphicx}
\usepackage{multirow}
\usepackage{longtable}
\usepackage{rotating}

\shorttitle{Cause of gamma-ray QPO in a blazar}
\shortauthors{Yan et al.}

\begin{document}

\title{Testing relativistic boost as the cause of gamma-ray quasi-periodic oscillation in a blazar}

\author{Dahai Yan}
\affil{Key Laboratory for the Structure and Evolution of Celestial Objects, Yunnan Observatory, Chinese Academy of Sciences, Kunming 650011, China; yandahai@ynao.ac.cn}
\affil{Center for Astronomical Mega-Science, Chinese Academy of Sciences, 20A Datun Road, Chaoyang District, Beijing 100012, China}

\author{Jianeng Zhou}
\affil{Shanghai Astronomical Observatory, Chinese Academy of Sciences, 80 Nandan Road, Shanghai 200030, China; zjn@shao.ac.cn}

\author{Pengfei Zhang}
\affil{Key Laboratory of Dark Matter and Space Astronomy, Purple Mountain Observatory, Chinese Academy of Sciences, Nanjing 210008, China; zhangpengfee@pmo.ac.cn}
\affil{Key Laboratory of Astroparticle Physics of Yunnan Province, Yunnan University, Kunming 650091, China}



\author{Qianqian Zhu}
\affil{Department of General Studies, Nanchang Institute of Science \& Technology, Nanchang 330108, China}

\author{Jiancheng Wang}
\affil{Key Laboratory for the Structure and Evolution of Celestial Objects, Yunnan Observatory, Chinese Academy of Sciences, Kunming 650011, China; yandahai@ynao.ac.cn}
\affil{Center for Astronomical Mega-Science, Chinese Academy of Sciences, 20A Datun Road, Chaoyang District, Beijing 100012, China}



\begin{abstract}

The mechanism for producing gamma-ray quasi-periodic oscillation (QPO) in blazar is unknown.
One possibility is the geometric model, in which without the need for intrinsic quasi-periodic variation, 
the relativistic Doppler factor changes periodically, resulting in observed gamma-ray QPO.
We propose a method to test this geometric model.
We analyze the {\it Fermi}-LAT data of PG 1553+113 spanning from 2008 August until 2018 February.
According to 29 four-month average spectral energy distributions (SEDs) in the energy range of 0.1-300 GeV, 
we split the {\it Fermi}-LAT energy range into three bands: 0.1-1 GeV, 1-10 GeV, and 10-300 GeV.
The spectrum in each energy range can be successfully fitted by a power-law.
The light curves and photon indices in the three energy ranges are obtained.
Then, light curves in three narrow energy ranges, i.e., 0.2-0.5 GeV, 2-5 GeV and 20- 40 GeV, are constructed, and 
the relative variability amplitudes in the three narrow energy ranges are calculated.
A discrete-correlation analysis is performed for the light curves.
Our results indicate that 
(i) the light curves in the different energy ranges follow the same pattern showed in the light curve above 0.1 GeV;
(ii) the three groups of photon indices in the energy ranges of 0.1-1 GeV, 1-10 GeV, and 10-300 GeV keep nearly constant;
(iii) the ratio between relative variability amplitudes in different narrow energy ranges are equal (within their errors) to the prediction by the Doppler effect. 
Our results support the scenario of the relativistic boost producing the gamma-ray QPO for PG 1553+113.

\end{abstract}



\section{Introduction} \label{sec:intro}

Blazars are the subclass of radio-loud active galactic nuclei (AGNs) with their relativistic jet pointing toward us \citep{urry}.  
Multi-wavelength radiations covering from MHz, optical to TeV gamma-ray energies have been observed from blazars.
Blazar emission is dominated by non-thermal radiation from the relativistic jet, therefore undergoing Doppler boosting.
The Doppler effect leads to flux enhancement and contraction of the variability timescales.

The Large Area Telescope (LAT) on the {\it Fermi} Gamma-ray
Space Telescope is providing continuous monitoring of the gamma-ray sky.
In the analysis of the LAT data of PKS 2155-304 spanning from 2008 August until 2014 June, 
\citet{14Sandrinelli} found a possible QPO signal with the $\sim1.7$ year period cycle.
Latter, \citet{Zhanga} analyzed the LAT data of PKS 2155-304 spanning from 2008 August until 2016 October, 
and found that this QPO signal is strengthened, with the significance of $\sim4.9\ \sigma$.
Taking advantage of LAT data, \citet{Ackermann} reported the gamma-ray QPO in PG 1553+113.
This signal is found in the LAT data covering from 2008 August to 2015 July with a 2.18$\pm$0.08 year period cycle.
Very recently, \citet{Tavani} confirmed the QPO signal in PG 1553+113 by updating the LAT data to 2017 September.

Possible gamma-ray QPOs with year-like timescales have been claimed in other several blazars \citep[e.g.,][]{16aSandrinelli,16bSandrinelli,Prokhorov,Zhangb,Zhangc} and 
in one narrow-line Seyfert 1 galaxy \citep{ZhangJin}.
The mechanism causing gamma-ray QPO in blazar remains unknown.
A few possibilities have been proposed \citep[e.g.,][]{Ackermann,Sobacchi,Caproni}, like pulsational accretion flow instabilities and jet precession.
The possible models can be divided into two classes: intrinsic origin and apparent origin.
The intrinsic origin refers to the case that QPO also exists in the comoving frame of the relativistic jet.
While in the scenario of the apparent origin, the intrinsic variation does not present QPO, and QPO is caused by a periodically changing Doppler factor.

Here we propose a method to test whether the gamma-ray QPO in a blazar is resulted from a periodically changing Doppler factor.
In Section~\ref{2} we describe our method, and results are presented in Section~\ref{3}; In Section~\ref{4} we will give a brief discussion.

\section{Method}
\label{2}
The Doppler factor is 
\begin{equation}
\delta_{\rm D}=\frac{1}{\Gamma\lbrack1-v/c\cdot \rm cos\theta\rbrack}\ ,
\end{equation}
where $\Gamma=(1-v^2/c^2)^{-1/2}$ is the bulk Lorentz factor, $v$ is the jet velocity, and $v\cdot \rm cos\theta$ is its line-of-sight component.
One can infer that the periodic change of the viewing angle $\theta$ results in a periodic variation of $\rm \delta_{\rm D}$.

Assuming that gamma-rays originate from a relativistically moving blob,  if the emission in the frame of the blob is isotropic, and follows a power-law distribution of the form $F'_{\nu'}\propto\nu'^{-\alpha}$, 
the flux density in the observer frame, $F_{\nu}$, is written as {\citep[e.g.,][]{urry} \footnote{\citet{Dermer} has carefully discussed the beaming pattern for such a blob. }
\begin{equation}
F_{\nu}(\nu)=\delta_{\rm D}^{3+\alpha}F'_{\nu'}(\nu)\ ;
\end{equation}
$\delta_{\rm D}^{\alpha}$ is the ratio of the intrinsic power-law fluxes at the observed and emitted frequencies (i.e., $\nu$ and $\nu'$), and $\nu=\delta_{\rm D}\nu'$.
Even if the intrinsic flux is constant, the periodic time modulation in $\theta$ will produce a periodic variation in the observed flux.

The relative variability amplitude is derived as \citep[e.g.,][]{DOrazio,Charisi}
\begin{equation}
A=\frac{\Delta F_{\nu}}{F_{\nu}}\propto \frac{3+\alpha}{\delta_{\rm D}}\ ,
\label{amp}
\end{equation}
where $\Delta F_{\nu}$ is the differential result of $F_{\nu}$ for time.
One can see that besides $\delta_{\rm D}$, the relative variability amplitude is also controlled by the spectral index $\alpha$.
The ratio between the relative variability amplitudes in low and high gamma-ray energies is 
\begin{equation}
\frac{A_{\rm L}}{A_{\rm H}}=\frac{3+\alpha_{\rm L}}{3+\alpha_{\rm H}}.
\label{ratio}
\end{equation}

Through testing whether the left and right sides of Equation~(\ref{ratio}) are consistent with each other, 
one can detect evidence for the Doppler effect in data.
Such a method has been applied to quasars to test the Doppler effect in their optical and UV data \citep{DOrazio,Charisi}.
These quasars are supermassive black hole binary candidates, and two black holes are separated by sub-pc, orbiting with mildly
relativistic velocities  \citep[e.g.,][]{DOrazio,Charisi}.

Considering the broad energy range of {\it Fermi}-LAT,  we can use Equation~(\ref{ratio})  to 
test whether the gamma-ray QPO in blazar is caused by Doppler boost.
For instance, we first analyze low and high energies gamma-ray light curves to determine the ratio of the observed variability
amplitudes, i.e. the left side of Equation~(\ref{ratio}). 
Next, we calculate the spectral indices in the two energy bands, and compute the ratio between the two indices, i.e. the right side of Equation~(\ref{ratio}).
For {\it Fermi}-LAT data, the two sides of Equation~(\ref{ratio}) can be obtained independently.
We then check whether the above two ratios are consistent within their errors.

In the following, the relative variability amplitude is defined as \citep[e.g.,][]{Nandra,abdo}
\begin{equation}
A=\sigma^2=\frac{S^2-\left \langle\sigma_{\rm err}^2 \right \rangle}{\left \langle F_{i} \right \rangle^2}\ ,
\label{sigma}
\end{equation}
where $S^2$ is the variance of the light curve \citep[e.g.,][]{Vaughan} and $\sigma_{\rm err}^2=\sigma_{i}^2+\sigma_{\rm sys}^2$ 
where $\sigma_{i}$ is the statistic error for flux $F_{i}$.
Following \cite{abdo}, we here use the systematic error $\sigma_{\rm sys}=0.03\langle F_{i}\rangle$. The error in $\sigma^2$ is evaluated by using the formula 
given in \citet{Edelson} and \cite{Vaughan}.

\section{Data analysis and results}
\label{3}

Taking PG 1553+113 as an example, we analyze its LAT data covering from 2008 August to 2018 February.

\subsection{Fermi-LAT data analysis}

We select $\sim 9.5$ years (spanning from MJD 542682.65 to MJD 58158.00) Fermi-LAT observations of PG 1553+113. 
The events of PASS 8 SOURCE class ({\tt evclass=128, evtype=3}), within a $20^{\circ} \times 20^{\circ}$ region of interest (ROI) centered at the position of PG 1553+113, are used \citep{2013arXiv1303.3514A}. We employ the standard {\tt ScienceTools} v10r0p5\footnote{https://fermi.gsfc.nasa.gov/ssc/data/analysis/software/} package for the analysis, 
and the {\tt P8R2\_SOURCE\_V6} instrument response function (IRF) is adopted. 
We exclude the events within zenith angles $> 90^{\circ}$ to minimize the effect from Earth limb. 
The 100 MeV - 300 GeV events are modeled by considering the target, the point sources within the ROI in 3FGL catalogue \citep{2015ApJS..218...23A}, and Galactic as well as extragalactic diffuse components respectively modeled by the two files {\tt gll\_iem\_v06.fits} and {\tt iso\_P8R2\_SOURCE\_V6\_v06.txt}. 
A binned maximum likelihood is performed to fit the events. 
First, model parameters for sources within $4^{\circ}$ of the center of the ROI are set free, 
while for sources outside $4^{\circ}$ but within $10^{\circ}$, only flux normalization is set free, and the sources outside $10^{\circ}$ have their parameters fixed 
to the values in 3FGL catalogue. 
Then, the likelihood analysis is ran to obtain best-fitting results.
Second, parameters of all sources except for our source of interest and variable sources (variability index $>72.44$) are fixed to the best-fitting values obtained in the first step, 
then we run the likelihood analysis to obtain best-fitting results in each time bin. 

\subsection{Results}

Following the above procedure, 
we first construct the four-month binned light curve in the energy range of 0.1-300 GeV (Fig.~\ref{alllc}), 
and also construct the SED in each time bin (Fig.~\ref{sed}).

The pattern of the light curve in Fig.~\ref{alllc} is similar to that reported in \citet{Tavani}.
The mentioned five peaks and 4.5 period cycles in \citet{Tavani} also can be found in Fig.~\ref{alllc}.
One can find more detailed discussions on this light curve in \citet{Tavani}. 

In Fig.~\ref{sed}, it seems that 
in some stages/bins the spectrum is not a simple power-law distribution, 
and the spectrum appears to become harder at $\sim1\ $GeV 
and then become softer at $\sim10\ $GeV (for example the SED in stage/bin 12).
Accordingly, we then build four-month binned light curves in three energy bands, namely 0.1-1 GeV, 1-10 GeV and 10-300 GeV (Fig.~\ref{three}).
For each energy band, we fit the spectrum in each time bin with a simple power-law, and obtain the photon index.
The photon indices in each energy range almost keep constant (Fig.~\ref{three})\footnote{Note that the photon index between 0.1-1 GeV has slight variation after $\sim$ MJD 57000.}. 
The average photon index is respectively $\Gamma_1=1.63\pm0.27$ in the range of 0.1-1 GeV, $\Gamma_2=1.58\pm0.11$ in the range of 1-10 GeV, 
and $\Gamma_3=1.98\pm0.20$ in the range of 10-300 GeV.
Therefore, only the break at 10 GeV is marginally significant.

Next, we construct light curves in the three narrow bands of 0.2-0.5 GeV, 2-5 GeV and 20-40 GeV (Fig.~\ref{narrow}). 
The flux $F$ in such a narrow band can be considered as the product of the central energy of the band $E_0$ and the
differential flux at the central energy of the band $F(E_0)$, i.e., $F=E_0\cdot F(E_0)$.
Because $E_0$ is a constant, we directly use $F$  to calculate the relative variability amplitudes in the three narrow bands with Equation~(\ref{sigma}).

In principle, we should use the photon indices in the narrow bands (e.g., 0.2-0.5 GeV) to calculate the right side of Equation~(\ref{ratio}). 
However, if we measure the photon index in such a narrow energy range, we could obtain a poorly constrained photon index. 
If we ensure that (1) a broader energy range includes the narrow energy range (e.g., 0.1-1 GeV including 0.2-0.5 GeV); 
and (2) the spectrum in the broader energy range (e.g., 0.1-1GeV) is a simple power-law, 
we could use a broader energy range (e.g., 0.1-1 GeV) to obtain the accurate photon index for the narrow energy range (e.g., 0.2-0.5 GeV).

The relative variability amplitude is respectively $A_1=0.046\pm0.021$ in the range of 0.2-0.5 GeV, $A_2=0.041\pm0.012$ in the range of 2-5 GeV, and $A_3=0.068\pm0.038$ in the range of 20-40  GeV.
We then obtain $\frac{A_1}{A_2}=1.12\pm0.61$, $\frac{A_2}{A_3}=0.61\pm0.39$, and $\frac{A_1}{A_3}=0.69\pm0.5$.
With the relation of photon index and spectral index $\Gamma=\alpha+1$ and using the values of $\Gamma_1$, $\Gamma_2$ as well as $\Gamma_3$, we have $\frac{3+\alpha_1}{3+\alpha_2}=1.02\pm0.08$, 
$\frac{3+\alpha_2}{3+\alpha_3}=0.9\pm0.05$, and $\frac{3+\alpha_1}{3+\alpha_3}=0.91\pm0.08$.
The results are plotted in Fig.~\ref{test}.
It seems that these values follow Equation~(\ref{ratio}) within their errors.

From the above analyses, we obtained seven light curves (see Figs.~\ref{alllc}, \ref{three} and \ref{narrow}).
In order to measure the correlation between the light curve in the energy band of 0.1-300 GeV and the other six light curves in different energy bands, 
we calculate their discrete correlation functions \citep[DCF;][]{Edelson98}.
The significance of correlation is calculated by using the method in \cite{dcf15} \citep[also see][]{dcf14}.
The results are showed in Fig.~\ref{dcf}.
One can see five peaks in each discrete-correlation result.
One peak is at the lag of 0, and the other four peaks are respectively at the lags of $\sim$850 days ($\sim2.3\ $year), $\sim$1700 days, $\sim$-850 days and $\sim$-1700 days.
The significances of the peaks are at the confidence levels of $\sim2\sigma$-$3\sigma$.
The results indicate that the light curves in the six different energy ranges have the same pattern with the light curve of above 0.1 GeV.

\begin{figure*}
 \centering
     \includegraphics[scale=0.6]{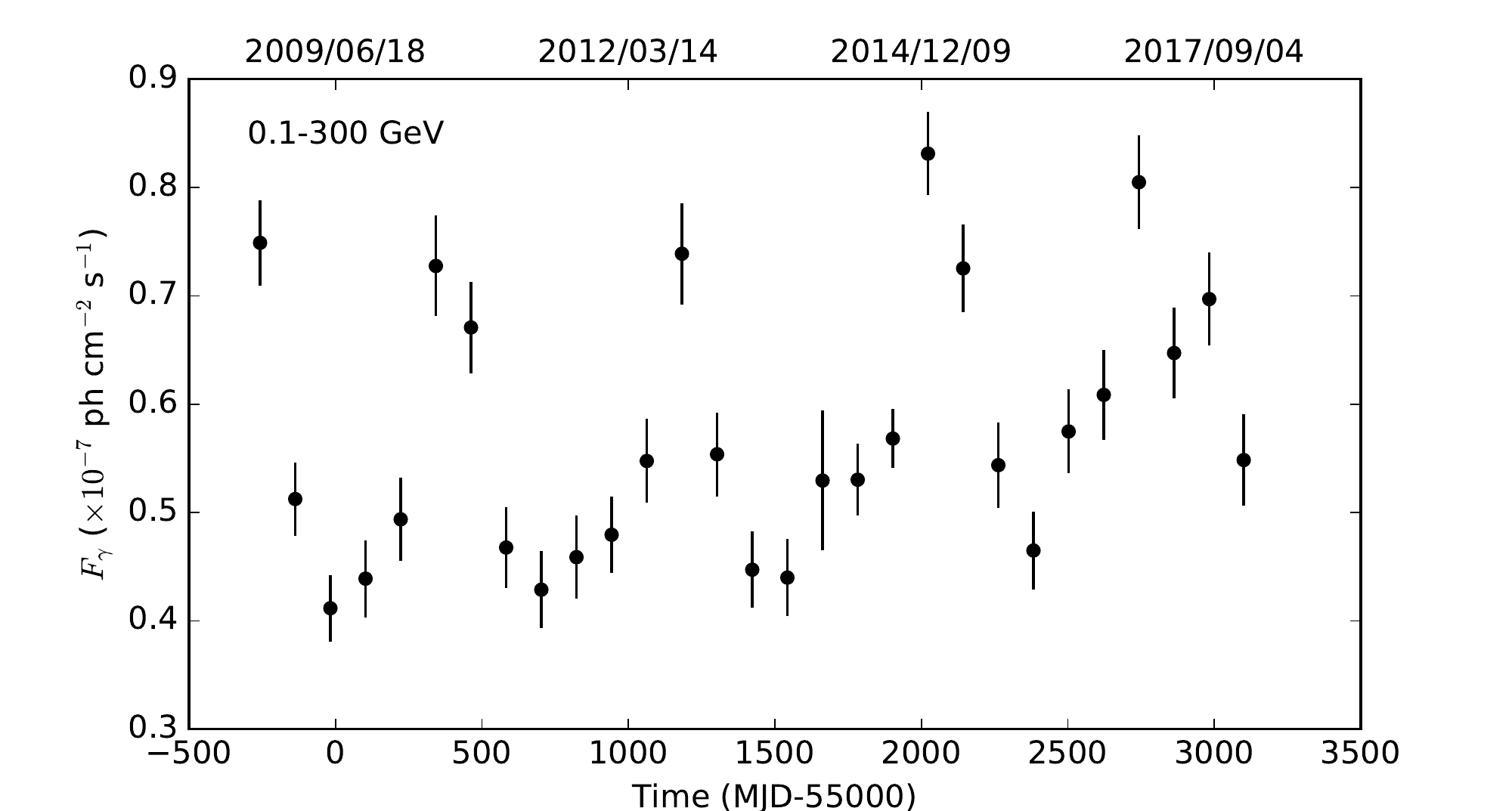}
\caption{PG 1553+113 gamma-ray light curve in the energy range of 0.1-300 GeV.} 
\label{alllc}
\end{figure*}

\begin{figure*}
   \centering
     \includegraphics[scale=0.4]{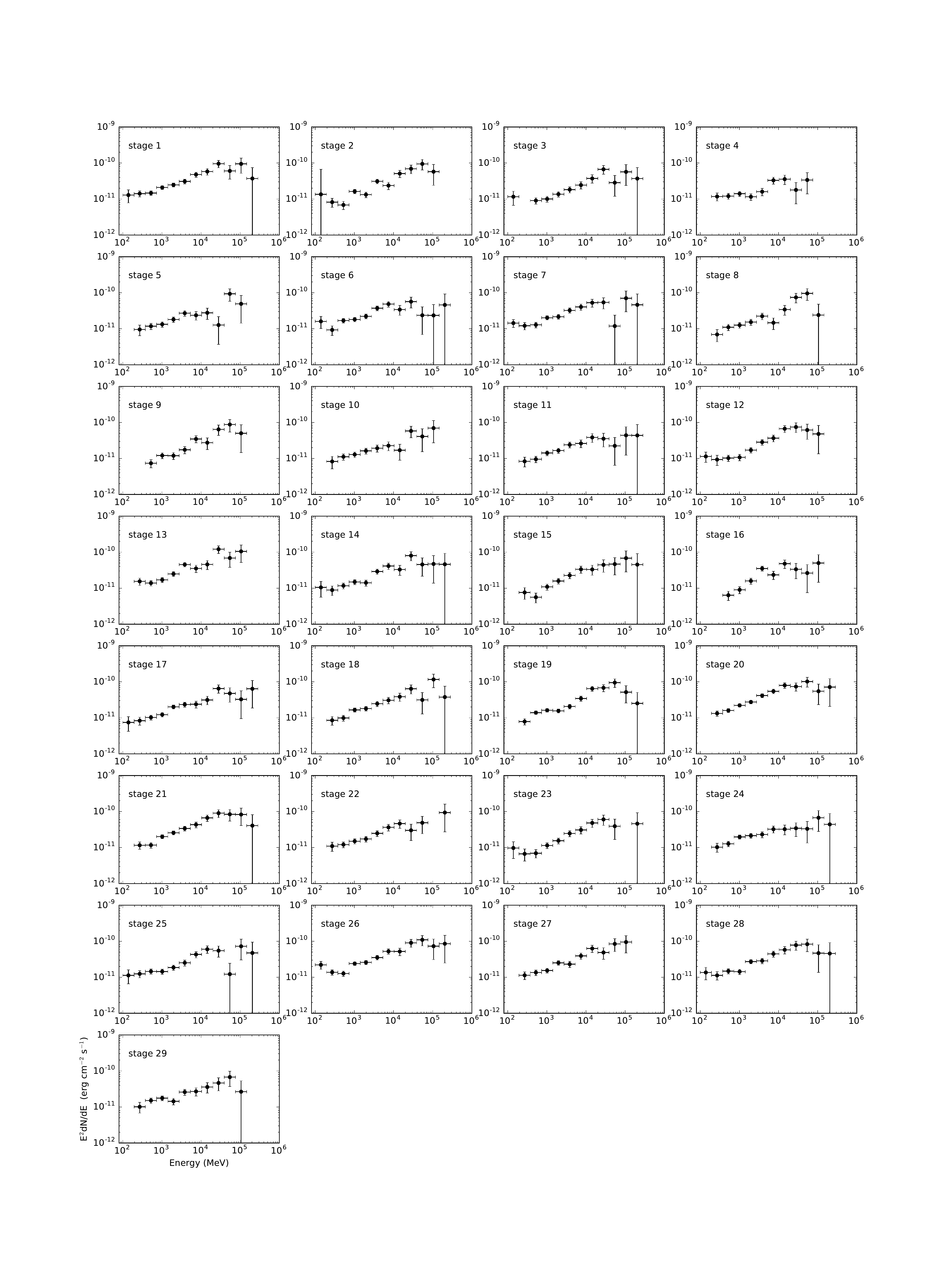}
\caption{SED in the energy range of 0.1-300 GeV in each time bin in Fig.~\ref{alllc}.} 
\label{sed}
\end{figure*}

\begin{figure*}
 \centering
     \includegraphics[scale=0.6]{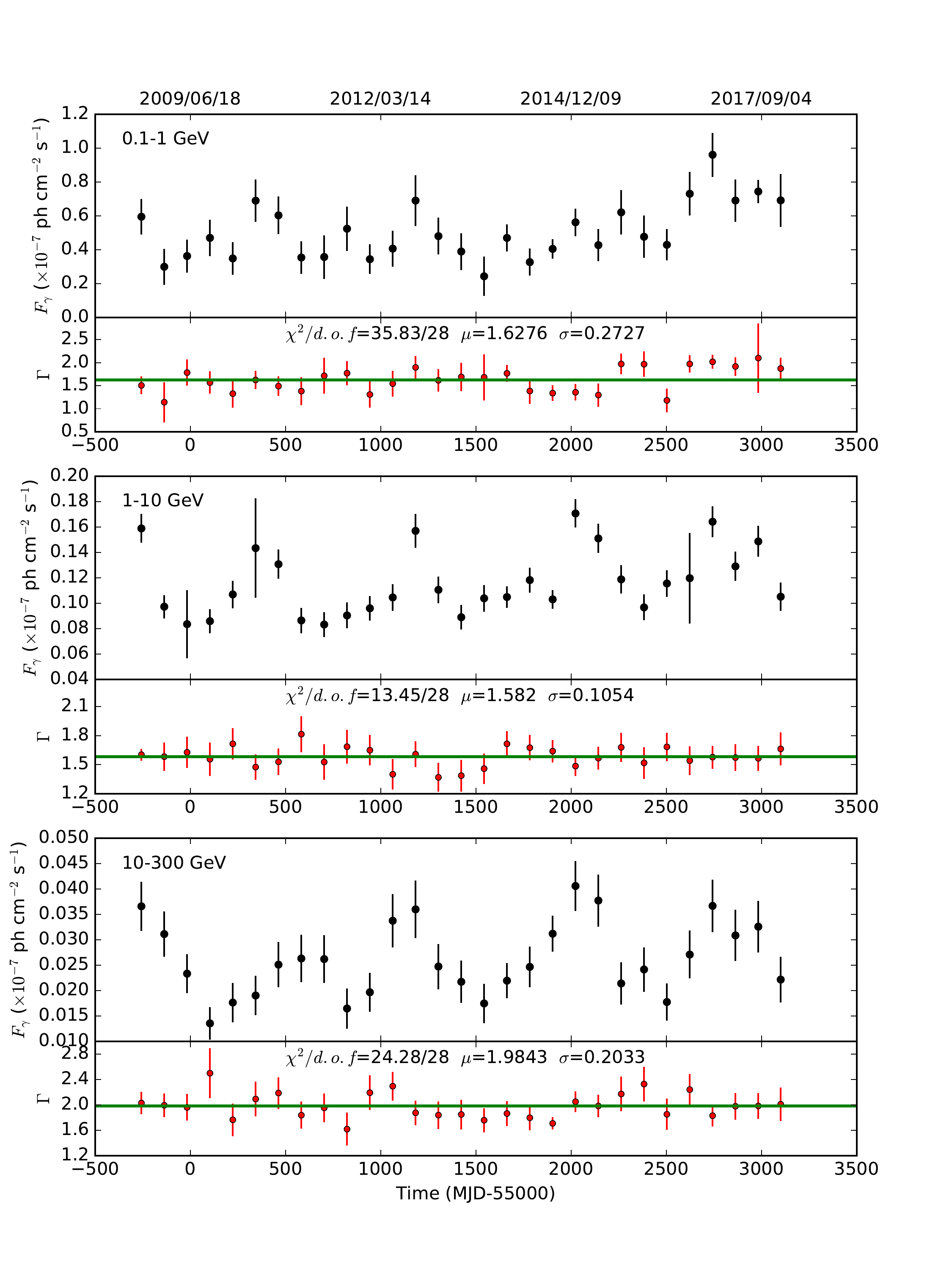}
\caption{From top to bottom: PG 1553+113 gamma-ray light curves and photon indices in the three energy bands of 0.1-1GeV, 1-10 GeV and 10-300 GeV. 
The solid line in each panel is the fitting result to the data with a constant.} \label{three}
\end{figure*}

\begin{figure*}
 \centering
     \includegraphics[scale=0.6]{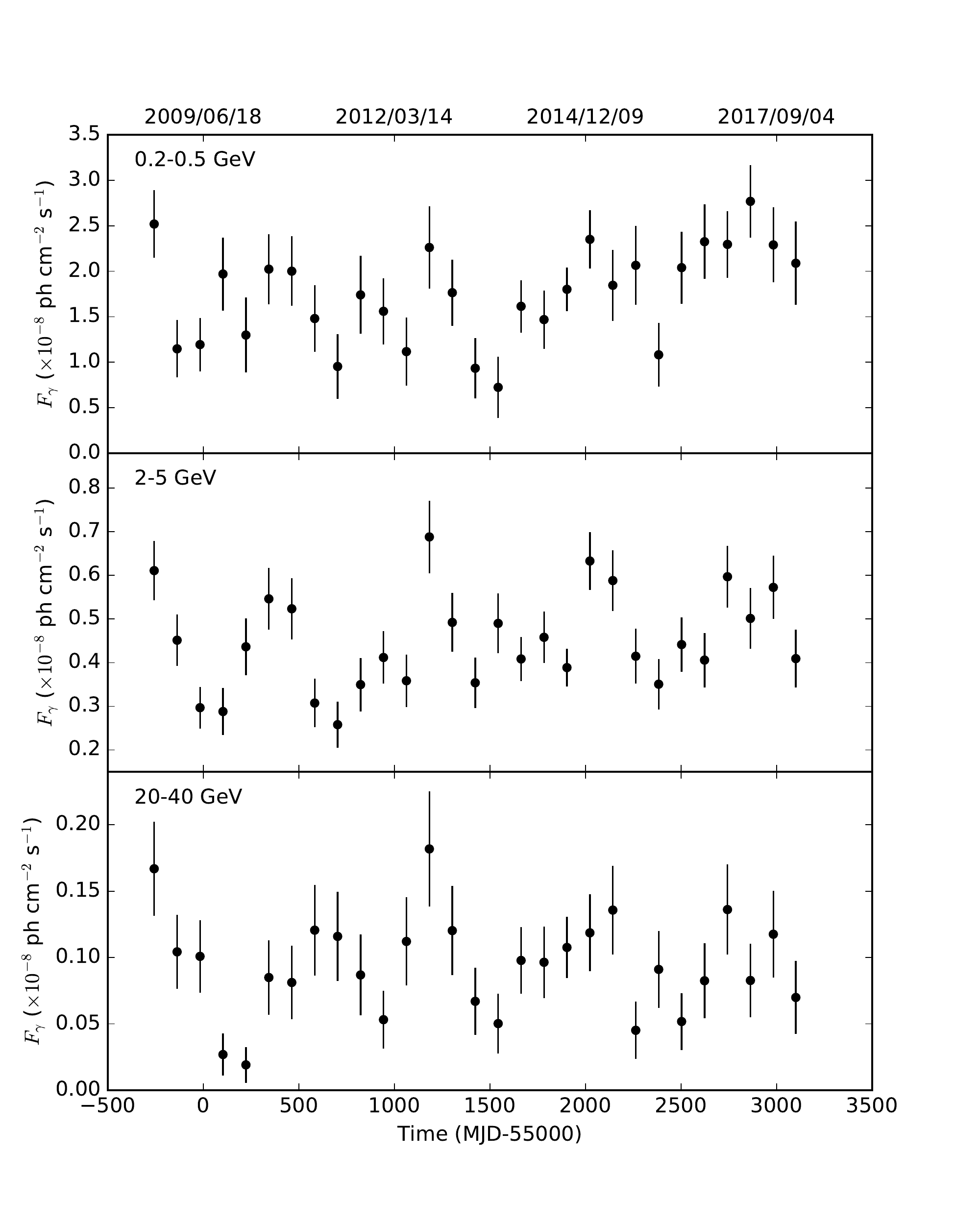}
\caption{From top to bottom: PG 1553+113 gamma-ray light curves in the three narrow energy bands of 0.2-0.5 GeV, 2-5 GeV and 20-40 GeV.} \label{narrow}
\end{figure*}

\begin{figure*}
 \centering
     \includegraphics[scale=0.6]{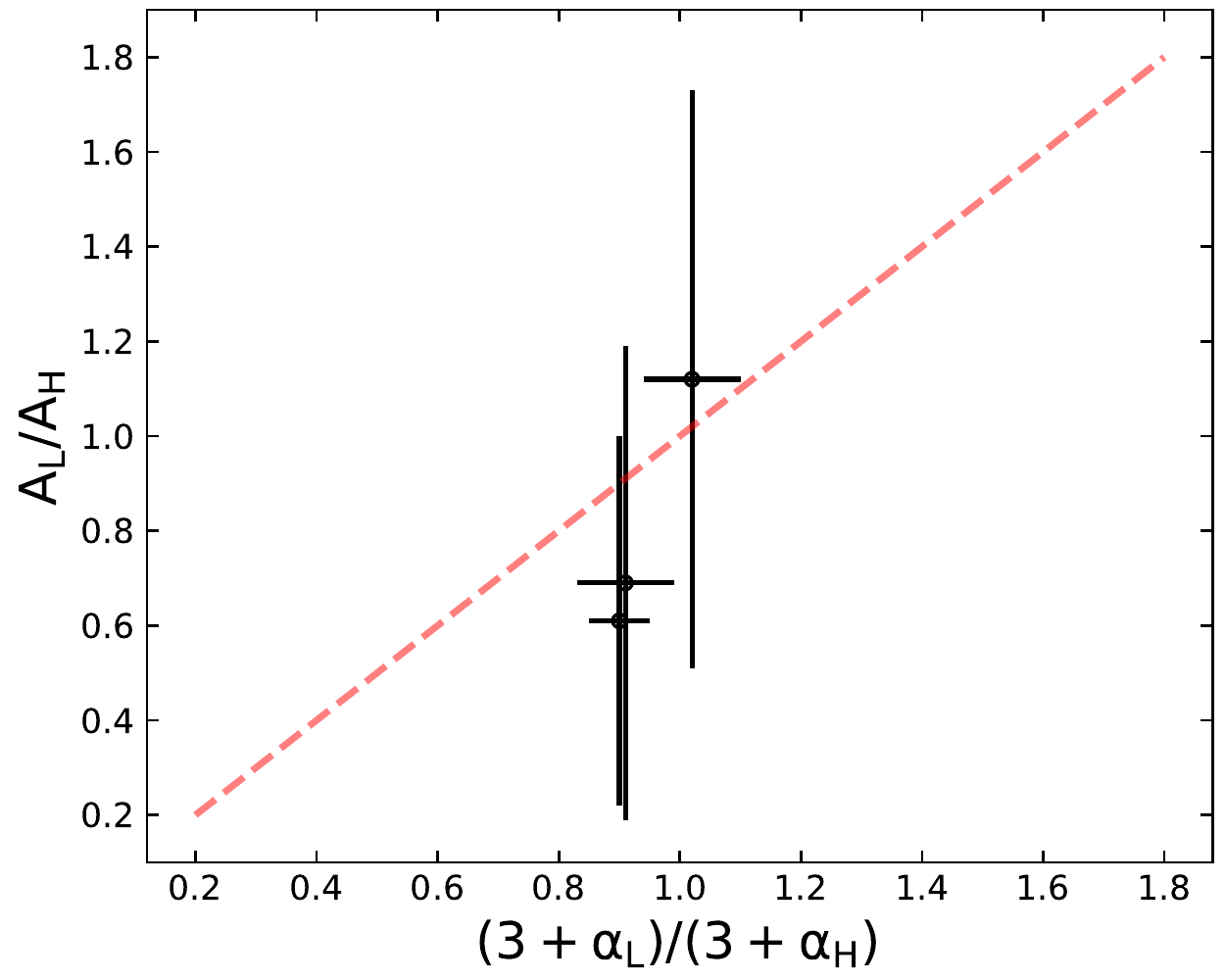}
\caption{The ratio of relative variability amplitudes, measured from the light curves 
versus the ratio of amplitudes expected from Doppler boosting, calculated
from the spectral indices. The dashed line represents $y=x$.}
\label{test}
\end{figure*}

\begin{figure*}
 \centering
     \includegraphics[scale=0.6]{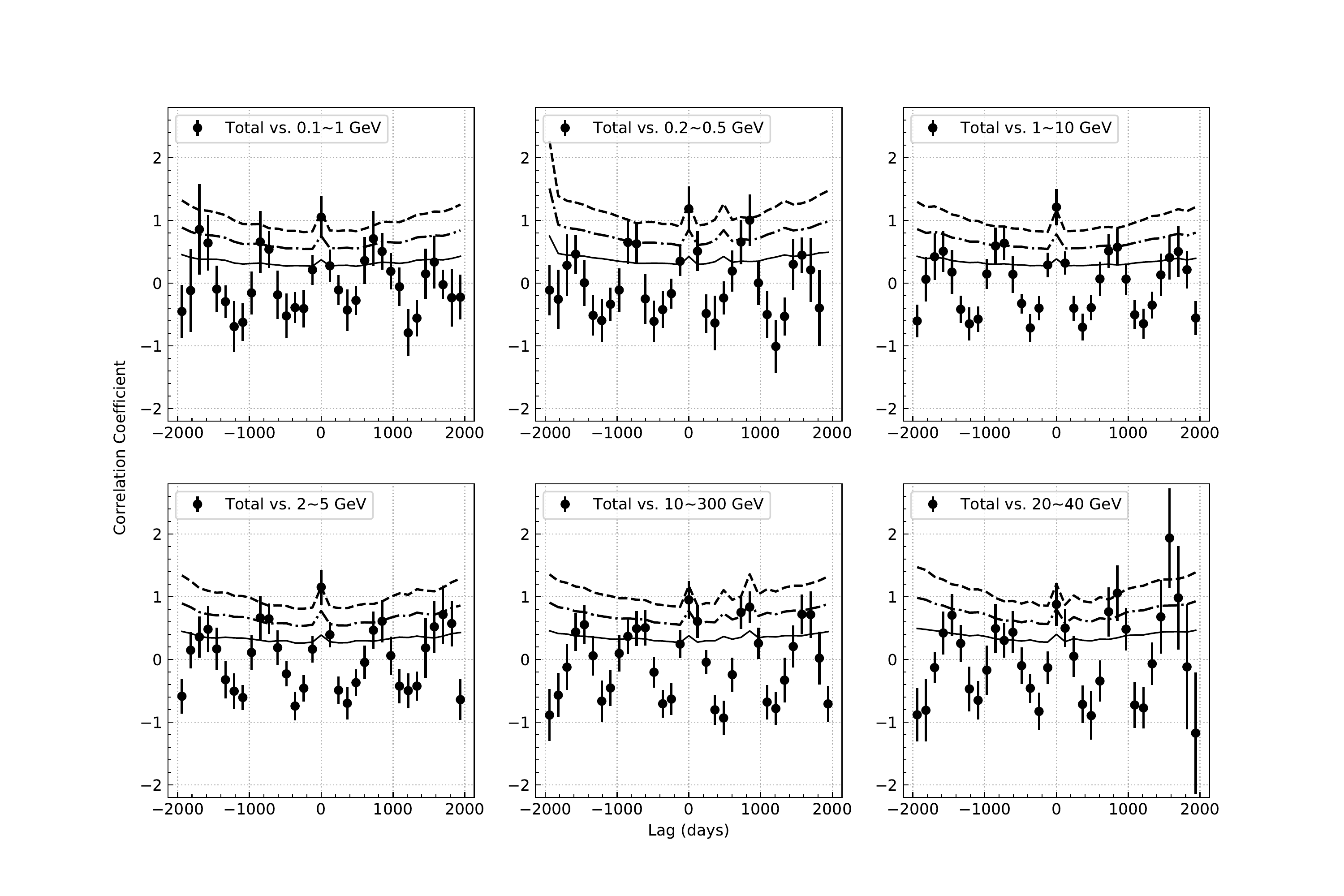}
\caption{DCF results of light curves in different energy ranges.  
The \emph{total} in each panel presents the energy range of 0.1-300 GeV. The points indicate discrete correlation coefficients. 
The dashed, dash-doted and solid lines respectively indicate the 3$\sigma$, 2$\sigma$ and 1$\sigma$ confidence levels. }
\label{dcf}
\end{figure*}

\section{DISCUSSION AND CONCLUSIONS}
\label{4}
We propose a model-independent approach to test whether the gamma-ray QPO in PG 1553+113 is only related to the relativistic jet itself.
According to the SEDs in the energy range of 0.1-300 GeV during different stages (Fig.~\ref{sed}), 
we split the LAT energy range into three energy bands, i.e., 0.1-1 GeV, 1-10 GeV and 10-300 GeV.
We construct light curves in the three energy bands, and obtain the photon indices in the three energy bands.
The three groups of photon indices all keep nearly constant in 10 years.
Three narrow energy bands are respectively selected in the above three energy ranges, and they are 0.2-0.5 GeV, 2-5 GeV and 20-40 GeV.
Light curves in the three narrow energy bands are built.

With the above observations we separately calculate the left and right side of Equation~(\ref{ratio}), and find that the values of the two sides 
are consistent with each other within the errors. This is in agreement with the prediction from the Doppler effect causing the gamma-ray QPO.
Namely the intrinsic emission is not necessary to be periodic, and a periodic modulation in Doppler factor causes the observed QPO.

Jet precession \citep[e.g.,][]{Romero,Caproni13}, helical structure \citep[e.g.,][]{VR,Ostorero,Raiteri09,Raiteri17} or rotation of a twisted jet \citep[e.g.,][]{VR,HP} could cause
emitting region to change its orientation and hence the Doppler factor. 
These models are called  geometrical models \citep{Rieger}.

For PG 1553+113, \citet{Raiteri} proposed an inhomogeneous
curved helical jet scenario to explain its complex UV to X-ray spectrum.
\citet{Caproni} identified seven distinct parsec-scale jet components in the jet of PG 1553+113, and found the evidence of jet precessing with a period of $\sim$2.2 years.
However, \citet{Caproni} argued that this jet precessing cannot account for the gamma-ray QPO in PG 1553+113, 
because of a delay  between the maxima of $\delta_{\rm D}$ and the periodic gamma-ray flares.
More works are needed to investigate the cause for a periodic modulation in Doppler factor.

It is noted that the photon index between 0.1-1 GeV has slight variation after $\sim$ MJD 57000 (Figs.~\ref{three} and \ref{twoM}).
The periodic light curve of 0.1-1 GeV after $\sim$ MJD 57000 is also slightly distorted.
An additional emission component below 1 GeV may be responsible for the distorted light curve and the variation of the photon index.
Looking at the epochs for the jet components in \citet{Caproni}, 
we suppose that the additional emission might be produced in the jet components of C6 and C7 through a shock-in-jet mechanism \citep[e.g.,][]{Marscher}.

In previous studies the gamma-ray QPOs were found in the integrated flux above 0.1 GeV or 1 GeV.
Here, we also detect similar QPO pattern in different energy bands (see Fig.~\ref{dcf}), 
and find that the QPO pattern above 1 GeV is more significant than that below 1 GeV.
\section*{Acknowledgements}
We thank the referee for the constructive questions. 
Wei Zeng and Shenbang Yang (YNU) are thanked for the help of calculating the DCF results.
We acknowledge financial supports from the National
Natural Science Foundation of China (NSFC-11803081, NSFC-11573060, NSFC-11573026, NSFC-U1738124, NSFC-11603059 and
NSFC-11661161010), and the Key Laboratory of Astroparticle
Physics of Yunnan Province (No. 2016DG006).
The work of D. H. Yan is also supported by the CAS
``Light of West China'' Program.
\bibliography{1553}

\appendix

In order to examine the impact of the bin size on our main results, we choose a 2-month bin to produce the results in Fig.~\ref{three}.
The results are shown in Fig.~\ref{twoM}. It is found that the bin sizes have little impact on our main results. A
possible cause for the slight distortion in the light curve of 0.1-1 GeV after $\sim$MJD 57000 is discussed in Section~\ref{4}. 

\begin{figure*}
 \centering
     \includegraphics[scale=0.56]{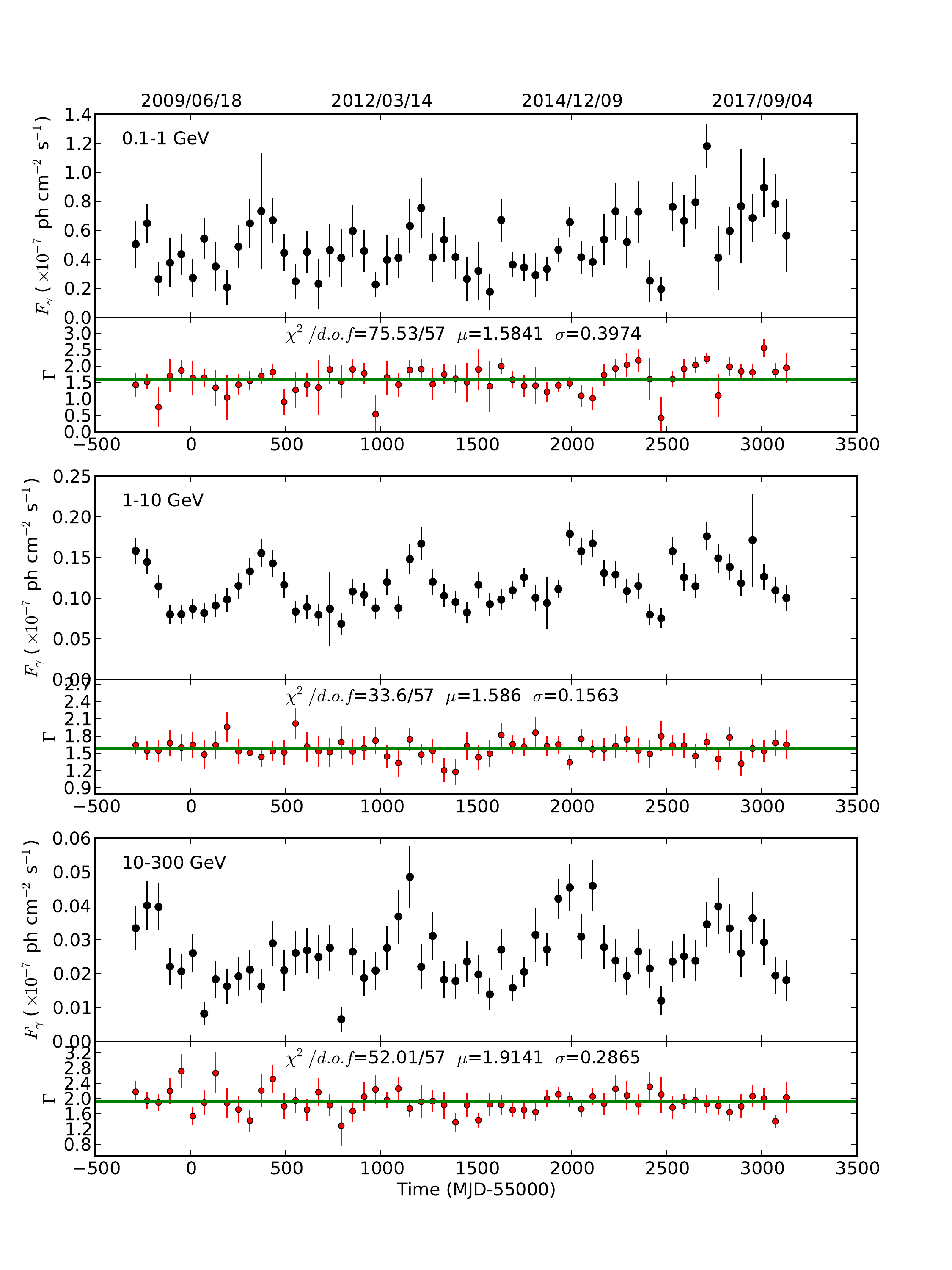}
\caption{Same as  Fig.~\ref{three}, but with the sampling bin of two months.} \label{twoM}
\end{figure*}



\end{document}